\documentstyle[aps,pra,epsfig,twocolumn]{revtex}

\def\be{\begin{equation}}
\def\ee{\end{equation}}
\def\bea{\begin{eqnarray}}
\def\eea{\end{eqnarray}}
\def\bma{\begin{mathletters}}
\def\ema{\end{mathletters}}
\def\C{\hbox{$\mit I$\kern-.7em$\mit C$}}
\def\c{{\cal C}}
\def\H{{\cal H}}
\def\E{{\cal E}}

\newcommand{\eins}{\mbox{$1 \hspace{-1.0mm}  {\bf l}$}}
\newcommand{\bra}[1]{\mbox{$\langle #1 |$}}
\newcommand{\ket}[1]{\mbox{$| #1 \rangle$}}
\newcommand{\braket}[2]{\mbox{$\langle #1  | #2 \rangle$}}
\newcommand{\proj}[1]{\ket{#1}\!\bra{#1}}

\begin{document}
\draft

\title{Three qubits can be entangled in two inequivalent ways}

\author{W. D\"ur, G. Vidal and J. I. Cirac}

\address{Institut f\"ur Theoretische Physik, Universit\"at Innsbruck,
A-6020 Innsbruck, Austria}

\date{\today}

\maketitle

\begin{abstract}
Invertible local transformations of a multipartite system are used to define
equivalence classes in the set of entangled states. This classification concerns
the entanglement properties of a single copy of the state. Accordingly, we say
that two states have the same kind of entanglement if both of them can be
obtained from the other by means of local operations and classical
communication (LOCC) with nonzero probability. When applied to pure states of a
three-qubit system, this approach reveals the existence of two inequivalent
kinds of genuine tripartite entanglement, for which the GHZ state and a W state
appear as remarkable representatives. In particular, we show that the W state
retains maximally bipartite entanglement when any one of the three qubits is
traced out. We generalize our results both to the case of higher dimensional
subsystems and also to more than three subsystems, for all of which we show
that, typically, two randomly chosen pure states cannot be converted into each
other by means of LOCC, not even with a small probability of success.
\end{abstract}


\narrowtext


\section{Introduction}

The understanding of entanglement is at the very heart of Quantum Information
Theory (QIT). In recent years, there has been an ongoing effort to characterize
qualitatively and quantitatively the entanglement properties of multiparticle
systems. A situation of particular interest in QIT consists of several parties
that are spatially separated from each other and share a composite system in an
entangled state. This setting conditionates the parties ---which are typically
allowed to communicate through a classical channel--- to only act locally on
their subsystems. But even restricted to local operations assisted with
classical communication (LOCC), the parties can still modify the entanglement
properties of the system and in particular they can try to convert an entangled
state into another. This possibility leads to natural ways of defining
equivalence relations in the set of entangled states ---where equivalent states
are then said to contain the same kind of entanglement---, also of establishing
hierarchies between the resulting classes.

For instance, we could agree in identifying any two states which can be obtained
from each other with certainty by means of LOCC. Clearly, this criterion is
interesting in QIT because the parties can use these two states indistintively
for exactly the same tasks. It is a celebrated result \cite{Be95} that, when
applied to many copies of a state, this criterion leads to identifying all
bipartite pure-state entanglement with that of the EPR state
$1/\sqrt{2}(\ket{00}+\ket{11}) $\cite{Ei35}. That is, the entanglement of any
pure state $\ket{\psi}_{AB}$ is asymptotically equivalent, under deterministic
LOCC, to that of the EPR state,  the entropy of entanglement $E(\psi_{AB})$
---the entropy of the reduced density matrix of either system $A$ or $B$---
quantifying the amount of EPR entanglement contained asymptotically in  $\ket{\psi}_{AB}$. In
contrast, recent contributions have shown that in systems shared by three or
more parties there are several inequivalent forms of entanglement under
asymptotic LOCC \cite{asy,Be99}.

This paper is essentially concerned with the entanglement
properties of a single copy of a state, and thus asymptotic
results do not apply here. For single copies it is known that
two pure states $\ket{\psi}$ and $\ket{\phi}$ can be obtained
with certainty from each other by means of LOCC if and only if
they are related by local unitaries LU \cite{Vi00J,Be99}. But
even in the simplest bipartite systems, $\ket{\psi}$ and
$\ket{\phi}$ are typically not related by LU, and continuous
parameters are needed to label all equivalence classes
\cite{Li97,Sc,Su00,Ca00,Ke99}. That is, one has to deal with
infinitely many kinds of entanglement. In this context an
alternative, simpler classification would be advisable.

One such classification is possible if we just demand that the conversion of the 
states is through stochastic local operations and classical communication 
(SLOCC) \cite{Be99}; that is, through LOCC but without imposing that it has to 
be achieved with certainty. In that case we can establish an equivalence 
relation stating that two states $\ket{\psi}$ and $\ket{\phi}$ are equivalent if 
the parties have a non-vanishing probability of success when trying to convert 
$\ket{\psi}$ into $\ket{\phi}$, and also $\ket{\phi}$ into $\ket{\psi}$ 
\cite{comment}. This relation has been termed stochastic equivalence in Ref.\ 
\cite{Be99}. Their equivalence under SLOCC indicates that both states are again suited 
to implement the same tasks of QIT, although this time the probability of a 
successful performance of the task may differ from $\ket{\phi}$ to $\ket{\psi}$. 
Notice in addition that since LU are a particular case of SLOCC, states 
equivalent under LU are also equivalent under SLOCC, the new classification 
being a coarse graining of the previous one.

The main aim of this work is to identify and characterize all possible kinds of
pure-state entanglement of three qubits under SLOCC. Unentangled states, and
also those which are product in one party while entangled with respect to the
remaining two, appear as expected, trivial cases. More surprising is the fact
that there are two different kinds of genuine tripartite entanglement. Indeed,
we will show that any (non-trivial) tripartite entangled state can be converted,
by means of SLOCC, into one of two standard forms, namely either the GHZ state \cite{Gr89}
\be
\ket{GHZ}=1/\sqrt{2}(|000\rangle+|111\rangle) \label{GHZ},
\ee
or else a second state
\be
|W\rangle=1/\sqrt{3}(|001\rangle+|010\rangle+|100\rangle) \label{W},
\ee
and that this splits the set of genuinely trifold entangled states into two sets
which are unrelated under LOCC. That is, we will see that if $\ket{\psi}$ can be
converted into the state $\ket{GHZ}$ in (\ref{GHZ}) and $\ket{\phi}$ can be
converted into the state $\ket{W}$ in (\ref{W}), then it is not possible to
transform, not even with only a very small probability of success, $\ket{\psi}$
into $\ket{\phi}$ nor the other way round.

The previous result is based on the fact that, unlike the GHZ state, not all
entangled states of three qubits can be expressed as a linear combination of
only two product states. Remarkably enough, the inequivalence under SLOCC of the
states $\ket{GHZ}$ and $\ket{W}$ can alternatively be shown from the fact that
the 3-tangle (residual tangle), a measure of tripartite correlations introduced
by Coffman et. al. \cite{Wo99}, does not increase on average under LOCC, as we
will prove here.

We will then move to the second main goal of this work, namely the analysis of
the state $\ket{W}$. It can not be obtained from a state $\ket{GHZ}$ by means of
LOCC and thus one could expect, in principle, that it has some interesting,
characteristic properties. Recall that in several aspects the GHZ state can be
regarded as the maximally entangled state of three qubits. However, if one of
the three qubits is traced out, the remaining state is completely unentangled.
Thus, the entanglement properties of the state $\ket{GHZ}$ are very fragile
under particle losses. We will prove that, oppositely, the entanglement of
$\ket{W}$ is maximally robust under disposal of any one of the three qubits, in
the sense that the remaining reduced density matrices\footnote{The reduced
density matrix $\rho_{AB}$ of a pure tripartite state $|\psi\rangle$ is defined
as $\rho_{AB}\equiv tr_C(|\psi\rangle\langle\psi|)$.} $\rho_{AB}$, $\rho_{BC}$
and $\rho_{AC}$ retain, according to several criteria, the greatest possible
amount of entanglement, compared to any other state of three qubits, either pure
or mixed.

We will finally analyze entanglement under SLOCC in more general multipartite
systems. We will show that, for most of these systems, there is typically no
chance at all to transform locally a given state into some other if they are
chosen randomly, because the space of entangled pure states depends on more
parameters than those that can be modified by acting locally on the subsystems.

The paper is organized as follows. In section II we characterize mathematically
the equivalence relation established by stochastic conversions under LOCC, and
illustrate its performance by applying it to the well-known bipartite case. In
section III we move to consider a system of three qubits, for which we prove the
existence of 6 classes of states under SLOCC ---including the 2 genuinely
tripartite ones---. Section IV is devoted to study the endurance of the entanglement of
the state $\ket{W}$ against particle losses. In section V more general multipartite systems are
considered. Section VI contains some
conclusions. Finally, appendix A to C prove, respectively, some needed results
related to SLOCC, the monotonicity of the 3-tangle under LOCC and the fact that
$\ket{W}$ retains optimally bipartite entanglement when one qubit is traced out.


\section{Kinds of entanglement under Stochastic LOCC}

In this work we define as equivalent the entanglement of two states $\ket{\psi}$
and $\ket{\phi}$ of a multipartite system iff local protocols exist that allow
the parties to convert each of the two states into the other one with some a
priori probability of success. In this approach, we follow the definition for
stochastic equivalence as given in \cite{Be99}\footnote{Stochastic transformations under LOCC had been
previously analyzed in \cite{Lo97,Vi99}.}. The underlying motivation for this
definition is that, if the entanglement of is $\ket{\psi}$ and $\ket{\phi}$ is
equivalent, then the two states can be used to perform the same tasks, although
the probability of a successful performance of the task may depend on the state
that is being used.

\subsection{Invertible local operators}

Sensible enough, this classification would remain useless if in practice we
would not be able to find out which states are related by SLOCC. Let us recall
that, all in all, no practical criterion is known so far that determines whether
a generic transformation can be implemented by means of LOCC. However, we can
think of any local protocol as a series of rounds of operations, where in each
round a given party manipulates locally its subsystem and communicates classically the
result of its operation (if it included a measurement) to the rest of parties.
Subsequent operations can be made dependent on previous results and the
protocol splits into several branches. This picture is useful because for our
purposes we need only focus on one of these branches. Suppose that state
$\ket{\psi}$ can be locally converted into state $\ket{\phi}$ with non-zero probability.
This means that at least one branch of the protocol does the job. Since we are
concerned only with pure states we can always characterize mathematically this
branch as an operator which factors out as the tensor product of a local
operator for each party. For instance, in a three-qubit case we would have that
$\ket{\psi}$ can be locally converted into $\ket{\phi}$ with some finite probability iff
an operator $A\otimes B \otimes C$ exists such that
\be
\ket{\phi} = A\otimes B \otimes C \ket{\psi},
\label{phipsi}
\ee
where operator $A$ contains contributions coming from any round in which party A
acted on its subsystem, and similarly for operators $B$ and $C$ \footnote{ In
practice the constraints $A^{\dagger}A,~B^{\dagger}B,~C^{\dagger}C \leq 1$
should be fulfilled if the invertible operators $A,B,C$ are to come from local
POVMs. In this work we do not normalize them in order to avoid introducing
unimportant constants to the equations. Instead, both the initial
and final states are normalized. }. Carrying on with the 3-qubit example, let us
now consider for simplicity that both states $\ket{\psi}$ and $\ket{\phi}$ have
rank 2 reduced density matrices $\rho_A \equiv $ tr$_{BC}(|\psi\rangle\langle\psi|), \rho_B$ and $\rho_C$. Then clearly the rank
of operators $A$, $B$ and $C$ need to be 2 (see appendix A). In other words,
each of these operators is necessarily invertible, and in particular
\be
\ket{\psi} = A^{-1}\otimes B^{-1}\otimes C^{-1} \ket{\phi}.
\ee
We see thus that, under the assumption of maximal rank for the reduced density
matrices, two-way convertibility implies the existence of invertible operators
$A$, $B$ and $C$ as in (\ref{phipsi}) [actually,  one-way convertibility alone
has already implied that an invertible local operator (ILO) $A\otimes B\otimes C$ exists].
Obviously, the converse also holds, namely that if an ILO
$A\otimes B\otimes C$ exists then for each direction of the conversion a local
protocol can be build that succeeds with non-zero probability. As explained in
appendix A in detail,  we can get rid of the previous assumption on the ranks
and announce with generality,

{\bf Result:} States $\ket{\psi}$ and $\ket{\phi}$ are equivalent under
stochastic local operations and classical communication ---SLOCC--- iff an
invertible local operator ---ILO--- relating them [as in, for instance, equation
(\ref{phipsi})] exists.

\subsection{Bipartite entanglement under SLOCC}

What does this classification implies in the well-known case \cite{Lo97,Vi99,Ni99} of bipartite
systems? Since LU are included in SLOCC, we can take the Schmidt decomposition
of a pure state $\ket{\psi}\in\C^n\otimes\C^m$, $n\leq m$, as the starting point
for our analysis. Thus,
\be
\sum_{i=1}^{n_{\psi}} \sqrt{\lambda_i} \ket{i}\otimes\ket{i} = U_A\otimes U_B\ket{\psi}; ~~~ \lambda_i >0, ~n_{\psi} \leq n,\label{Schmidt1}
\ee
where $U_A$ and $U_B$ are some proper local unitaries, the coefficients
$\lambda_i$ decrease with $i$, and $n_{\psi}$ is the number of non-vanishing
terms in the Schmidt decomposition. Clearly, the ILO
\be
\frac{1}{\sqrt{n_\psi}}(\sum_{i=1}^{n_{\psi}} \frac{1}{\sqrt{\lambda_i}}\proj{i} + \sum_{i=n_{\psi}+1}^n \proj{i})\otimes 1_B
\ee
transforms (\ref{Schmidt1}) into a maximally entangled state
\be
\frac{1}{\sqrt{n_{\psi}}}\sum_i^{n_{\psi}} \ket{i}\otimes\ket{i},
\label{maxi}
\ee
which depends only on the Schmidt number $n_{\psi}$. Since SLOCC cannot modify
the rank of the reduced density matrices $\rho_A$ and $\rho_B$, which is given
by $n_{\psi}$, we conclude that in $\C^n\otimes\C^m$, $n\leq m$, there are $n$
different kinds of entangled states, corresponding to $n$ different classes
under SLOCC.  Each of these classes is characterized by a given Schmidt number,
and we can choose as their representatives the state (\ref{maxi}) with
$n_{\psi}=1,...,n$. Clearly $n_{\psi}=1$ corresponds to states that are less
entangled than the rest (they are, after all, unentangled). This hierarchical
relation can be extended to the rest of classes by noting that none-invertible
local operators can project out some of the Schmidt terms and thus diminish the
Schmidt number of a state. Therefore the state $\ket{\psi}$ can be locally
converted into the state $\ket{\phi}$ with some finite probability iff
$n_{\psi}\geq n_{\phi}$, or in terms of kinds of entanglement, we can say that
the entanglement of the class characterized by a given Schmidt number is more
powerful than that of a class with a smaller Schmidt number.

For later reference we also note that in a two-qubit system,
$\H=\C^2\otimes\C^2$, we can write any state, after using a convenient LU,
uniquely as
\be
\ket{\psi} = c_{\delta}~ \ket{0}\otimes\ket{0} + s_{\delta} ~\ket{1}\otimes\ket{1}; ~~~~~c_{\delta} \geq s_{\delta}\geq 0,
\ee
where $c_{\delta}, s_{\delta}$ stand for cos$\delta$ and sin$\delta$. This is
either a product (unentangled) state $\ket{\psi_{A-B}}=\ket{0}\otimes \ket{0}$
for $c_{\delta}=1$ or else an entangled state that can be converted into the EPR
state,
\be
\frac{1}{\sqrt{2}}(\ket{0}\otimes\ket{0}+\ket{1}\otimes\ket{1}),
\label{EPR}
\ee
with probability $p=E_{2}(\psi)$, where $E_2(\psi)\equiv \lambda_2$ is the
entanglement monotone that provides a quantitative description of the non-local
resources contained in a single copy of a two-qubit pure state \cite{monotones}.
Any state $\ket{\psi}$ can be obtained from state (\ref{EPR}) with certainty,
this contributing to the fact that the EPR state is considered the maximally
entangled state of two qubits.



\section{Entanglement of pure states of three qubits}

In this section we analyze a system of three qubits. We show that SLOCC
split the set of pure states into 6 inequivalent classes, which further
structure themselves into a three-grade hierarchy when non-invertible local
operations are used to relate them. At the top of the hierarchy we find two
inequivalent classes of true tripartite entanglement, which we name GHZ-class
and W-class after our choice of corresponding representatives. The three
possible classes of bipartite entanglement are accessible (with some
non-vanishing probability) from {\em any} state of the W and GHZ classes by
means of a non-invertible local operator. Finally, at the bottom of the
hierarchy we find non-entangled states.

The ranks r$(\rho_A)$, r$(\rho_B)$ and r$(\rho_C)$ of the reduced density
matrices, together with the range $R(\rho_{BC})$ of $\rho_{BC}$, will be the
main mathematical tools used through the first part of this section. By
analysing them we will be able to make an exhaustive classification of
three-qubit entanglement. Later on we will rephrase some of these results in
terms of well-known measures of entanglement. In particular, we will see that
the existence of two inequivalent kinds of true tripartite entanglement under
SLOCC is very much related to the fact that the 3-tangle, a measure of
tripartite entanglement introduced in \cite{Wo99}, is an entanglement monotone
(see appendix B).

At the end of the section also a practical way to identify the
class an arbitrary state belongs to will be discussed.

\subsection{Non-entangled states and bipartite entanglement.}

If at least one of the local ranks r$(\rho_A)$, r$(\rho_B)$ or r$(\rho_C)$ is 1,
then the pure state of the three qubits factors out as the tensor product of two
pure states, and this implies that at least one of the qubits is uncorrelated
with the other two. SLOCC distinguish states with this feature depending on which
qubits are uncorrelated from the rest.

\vspace{.5cm}

\noindent {\bf Class A-B-C (product states)} \\
This class corresponds to states with $r(\rho_A)=r(\rho_B)=r(\rho_C)=1$. They
can be taken, after using some convenient LU, into the form
\be
\ket{\psi_{A-B-C}} = \ket{0}\ket{0}\ket{0},
\ee
where we have already relaxed the notation for
$\ket{0}\otimes\ket{0}\otimes\ket{0}$.

\vspace{.5cm}

\noindent {\bf Classes A-BC, AB-C and C-AB \\
(bipartite entanglement)} \\
These three classes of states contain only bipartite entanglement between two of
the qubits, one of the reduced density matrices having rank 1 and the other two
having rank 2. For example, the states in class $A-BC$ possess entanglement
between the systems $B$ and $C$ ($r(\rho_B)=r(\rho_C)=2$) and are product with
respect to system $A$ ($r(\rho_A)=1$). LU allow us to write uniquely states of
the class $A-BC$ as
\be
|\psi_{A-BC}\rangle=|0\rangle(c_\delta|0\rangle|0\rangle+s_\delta |1\rangle|1\rangle), ~~~ c_{\delta} \geq s_{\delta}>0,
\ee
and similary for $|\psi_{B-AC}\rangle$ and $|\psi_{C-AB}\rangle$.
We choose the maximally entangled state
\be
\frac{1}{\sqrt{2}} |0\rangle(|0\rangle|0\rangle+|1\rangle|1\rangle)
\label{repreA-BC}
\ee
as representative of the class $A-BC$. Any other state within this class can be
obtained from (\ref{repreA-BC}) with certainty by means of LOCC.

\vspace{.5cm}

The proof that these four marginal classes are inequivalent under SLOCC is very
simple. We only need to recall that the local ranks are invariant under ILO
(see appendix A). In what follows we will analyze the more interesting case of
$r(\rho_{\kappa})=2,~\kappa = A,B,C$. To see that there are two inequivalent
classes fulfilling this condition we will have to study possible product
decompositions of pure states.

\subsection{True three-qubit entanglement.}

There turns out to be a close connection between convertibility under SLOCC and
the way entangled states can be expressed minimally as a linear combination of
product states. For instance, as we shall prove later on, the GHZ and W states
have a different number of terms in their minimal product decompositions
(\ref{GHZ}) and (\ref{W}), namely 2 and 3 product terms respectively, and this
readily implies that there is no way to convert one state into the other by
means of an ILO $A\otimes B\otimes C$. Indeed, let us consider,
e.g., the most general pure state that can be obtained reversibly from a
$\ket{GHZ}$. It reads
\be
A\otimes B\otimes C\ket{GHZ}=\frac{1}{\sqrt{2}}( \ket{A0}\ket{B0}\ket{C0}+\ket{A1}\ket{B1}\ket{C1}),
\label{new}
\ee
where $\ket{A0}$ and $\ket{A1}$ are linearly independent vectors (since $A$ is
invertible) and similarly for the other two qubits. That is, the minimal number
of terms in a product decomposition for the state (\ref{new}) is also 2.
Actually, we have that also for a general multipartite system,

{\bf Observation:} The minimal number of product terms for any given state
remains unchanged under SLOCC.

This simple observation tells us already that in three qubits there are at least
two inequivalent kinds of genuine tripartite entanglement under SLOCC, that of
$\ket{GHZ}$ and that of $\ket{W}$.

However, we still have to prove that the state $\ket{W}$ cannot be expressed as
a linear combination of just two product vectors. In order to complete our
classification we also have to show that any pure state of three qubits with
maximal local ranks can be reversibly converted into either the state
$\ket{GHZ}$ or the state $\ket{W}$. We start with an obvious lemma regarding
product decompositions:

{\bf Lemma:} Let $\sum_{i=1}^l \ket{e_i}\ket{f_i}$ be a product decomposition
for the state $\ket{\eta}\in\H_{E}\otimes\H_{F}$. Then the set of states
$\{\ket{e_i}\}_{i=1}^l$ span the range of $\rho_E\equiv$Tr$_F \proj{\eta}$.

{\bf Proof:} We have that $\rho_E = \sum_{i,j=1}^l \braket{f_i}{f_j}
\ket{e_j}\bra{e_i}$. On the other hand $\ket{\nu}$ is in the range of $\rho_E$
iff a state $\ket{\mu}$ exists such that $\ket{\nu}=\rho_E\ket{\mu}$, that is
$\ket{\nu} = \sum_{i,j=1}^l \braket{f_i}{f_j}\braket{e_i}{\mu} \ket{e_j}$.
$\Box$

In particular, $r(\rho_A)=2$ implies that at least two product terms are needed
to expand $\ket{\psi}\in \C^2\otimes\C^2\otimes\C^2$. Let us suppose that a
product decomposition with only two terms is possible, namely
\be
\ket{\psi}=\ket{a_1}\ket{b_1}\ket{c_1} + \ket{a_2}\ket{b_2}\ket{c_2}.
\label{2deco}
\ee
Then, also according to the previous lemma, $\ket{b_1}\ket{c_1}$ and
$\ket{b_2}\ket{c_2}$ have to span the range of $\rho_{BC}$, R$(\rho_{BC})$.

But R$(\rho_{BC})$ is a two dimensional subspace of $\C^2\otimes\C^2$.
Therefore it always contains either only one or only two product states
\cite{STV} [unless R$(\rho_{BC})$ was supported in $\C\otimes\C^2$ or
$\C^2\otimes\C$, but we already excluded this possibility because we are
considering r$(\rho_B)=$r$(\rho_C)=2$]. Notice that a two-term decomposition
(\ref{2deco}) requires that R$(\rho_{BC})$ contains at least two product vectors.
Only one product vector in R$(\rho_{BC})$, and thus the impossibility of
decomposition (\ref{2deco}), is going to be precisely the trait of the
states in the W-class.

\vspace{.5cm}

\noindent {\bf GHZ-class}\\
Let us suppose first that R$(\rho_{BC})$ contains two product vectors,
$\ket{b_1}\ket{c_1}$ and $\ket{b_2}\ket{c_2}$. Then decomposition (\ref{2deco})
is possible, and actually unique, with $\ket{a_i}= \braket{\xi_i}{\psi}$,
$i=1,2$, where $\ket{\xi_i}$ are the two vectors supported in $R(\rho_{BC})$
that are biorthonormal to the $\ket{b_i}\ket{c_i}$. In this case we can use LU
in order to take $\ket{\psi}$ into the useful standard product form (see also \cite{Ac00})
\be
|\psi_{GHZ}\rangle=\sqrt{K}(c_\delta|0\rangle|0\rangle|0\rangle+s_\delta e^{i\varphi}|\varphi_A\rangle|\varphi_B\rangle|\varphi_C\rangle), \label{GHZclass}
\ee
where
\bea
\ket{\varphi_A}=c_{\alpha}\ket{0}+s_{\alpha}\ket{1}\nonumber\\
\ket{\varphi_B}=c_{\beta}\ket{0}+s_{\beta}\ket{1}\nonumber\\
\ket{\varphi_C}=c_{\gamma}\ket{0}+s_{\gamma}\ket{1}
\eea
and $K=(1+2 c_\delta s_\delta c_\alpha c_\beta c_\gamma c_\varphi)^{-1} \in
(1/2,\infty)$ is a normalization factor. The ranges for the five parameters are
$\delta \in (0,\pi/4], \alpha,\beta,\gamma \in (0,\pi/2]$ and $\varphi \in
[0,2\pi)$.

All these states are in the same equivalence class as the $\ket{GHZ}$
(\ref{GHZ}) under SLOCC. Indeed, the ILO
\bea
\sqrt{2K}\left( \begin{array}{ll} c_\delta & s_\delta c_\alpha e^{i\varphi}  \\0 & s_\delta s_\alpha e^{i \varphi} \end{array} \right) \otimes
\left( \begin{array}{ll} 1 & c_\beta \\0 & s_\beta \end{array} \right) \otimes
\left( \begin{array}{ll} 1 & c_\gamma\\0 & s_\gamma \end{array} \right),
\eea
applied to $\ket{GHZ}$ produces the state (\ref{GHZclass}).

The GHZ state is a remarkable representative of this class. It is maximally
entangled in several senses \cite{Gi98}. For instance, it maximally violates
Bell-type inequalities, the mutual information of measurement outcomes is
maximal, it is maximally stable against (white) noise and one can locally obtain
from a GHZ state with unit propability an EPR state shared between any two of
the three parties. Another relevant feature is that when any one of the three
qubits is traced out, the remaining two are in a separable ---and therefore
unentangled--- state.

\vspace{.5cm}

\noindent {\bf W-class}\\
Let us move to analyze the case where R$(\rho_{BC})$ contains only one product
vector. We already argued that decomposition (\ref{2deco}) is now not possible.
Instead we can (uniquely) write
\be
\ket{\psi} = \ket{a_1}\ket{b_1}\ket{c_1} + \ket{a_2}\ket{\phi_{BC}},
\label{const}
\ee
where $\ket{\phi_{BC}}$ is the vector of $R(\rho_{BC})$ which is orthogonal to
$\ket{b_1}\ket{c_1}$, and $\ket{a_1}$ and $\ket{a_2}$ are given by
$\braket{b_1|\langle c_1}{\psi}$ and $\braket{\phi_{BC}}{\psi}$. By means of LU
(\ref{const}) can be always rewritten as
\bea
\ket{\psi} = (\sqrt{c}\ket{1}+ \sqrt{d}\ket{0})&&\ket{00} \nonumber\\
+\ket{0}&&(\sqrt{a}\ket{01} + \sqrt{b}\ket{10}).
\label{Wclass2}
\eea
Indeed, we first take $\ket{b_1}\ket{c_1}$ into $\ket{0}\ket{0}$. Then, since
$\ket{\phi_{BC}}$ has been chosen orthogonal to $\ket{b_1}\ket{c_1}$, it must
become $x\ket{01} +y\ket{10}+z \ket{11}$. By requiring that a linear combination
of these two vectors has no second product vector we obtain that $z=0$
\cite{product}. In addittion the coefficients $ \sqrt{a}\equiv x,~\sqrt{b}\equiv
y,~\sqrt{c}$ and $\sqrt{d}$ can be made positive by absorbing the three relative
phases into the definition of state $\ket{1}$ of subsystems $A$, $B$ and $C$.
Thus case (i) has been taken into the form (\ref{Wclass2}) by just using LU.
Before we showed that 2 terms could not suffice in a product decomposition of
the state. Now we see that 3 product terms always do the job, for instance
$(\sqrt{c}\ket{1}+ \sqrt{d}\ket{0})\ket{00}$, $\sqrt{a}\ket{0}\ket{01}$ and
$\sqrt{b}\ket{0}\ket{10}$ once we took the original state into the standard,
unique form
\be
\ket{\psi_W} = \sqrt{a} \ket{001} +\sqrt{b} \ket{010} + \sqrt{c} \ket{100} +\sqrt{d} \ket{000},
\label{Wclass}
\ee
where $a, b, c >0$, and $d \equiv 1- (a + b+ c) \geq 0$.

The parties can locally obtain the state (\ref{Wclass}) from the state $\ket{W}$
in (\ref{W}), which we choose as a representative of the class ---and whose
study we postpone for later on---, by application of an ILO of the form
\bea
\left( \begin{array}{ll} \sqrt{a} & \sqrt{d} \\0 & \sqrt{c} \end{array}
\right) \otimes
\left( \begin{array}{ll} \sqrt{3} & 0 \\0 & \frac{\sqrt{3b}}{\sqrt{a}}
\end{array} \right) \otimes
\left( \begin{array}{ll} 1 & 0\\0 & 1 \end{array} \right).
\eea

\vspace{.5cm}

Before moving to relate these classes by means of non--invertible local operators, we note that
states within the GHZ-class and the W-class depend, respectively, on 5 and 3
parameters that cannot be changed by means of LU. Previous works
\cite{Li97,Sc,Ac00,Ca00} have shown that a generic state of three qubits
depends, up to LU, on 5 parameters. This means that states tipically belong to
the GHZ-class, or equivalently, that a {\em generic} pure state of three qubits
can be locally transformed into a GHZ with finite probability of success (see
also \cite{Co00}). The W-class is of zero measure compared to the GHZ-class.
This does not mean, however, that it is irrelevant. In a similar way as
separable mixed states are not of zero measure with respect to entangled states,
even though product states are, it is in principle conceivable that mixed states
having only W-class entanglement are also not of zero measure in the set of
mixed states.

\subsection{Relating SLOCC--classes by means of non--invertible operators}

In this subsection, we investigate the hierarchical relation of the 6 SLOCC-equivalence
classes under non--invertible operators, i.e. under general LOCC.

A non--invertible local operator transforms $\ket{\psi}$ into $\ket{\phi}$ according to
(\ref{phipsi}), but with at least one of the local operators $A$, $B$ and $C$
having rank 1. This means that the local ranks of the pure states can be
diminished. For instance, if the initial state $\ket{\psi}$ belongs either to
the GHZ or W class, then a non-invertible operator will diminish at least one of
the local ranks. That is, $\ket{\phi}$ belongs necessarily to one of the
bipartite classes  $\kappa-\mu\nu~ (\kappa\not=\mu\not=\nu \in\{A,B,C\})$ or
else is a product state $A-B-C$.

\begin{figure}[ht]
\begin{picture}(230,220)
\put(5,55){\epsfxsize=230pt\epsffile[36 695 232 832]{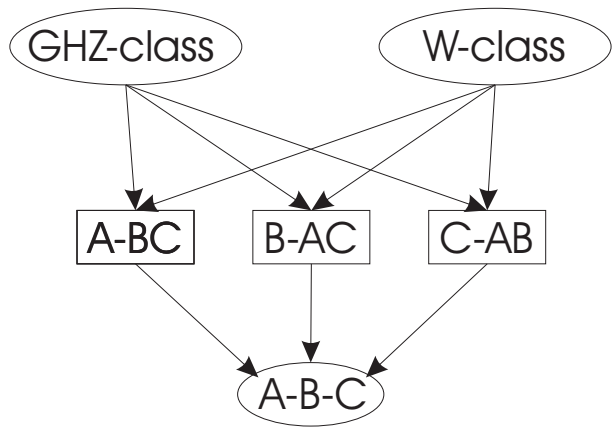}}
\end{picture}
\caption[]{Different local classes of tripartite pure states. The direction of the arrows indicate
which non--invertible transformations between classes are possible.}
\label{Fig1}
\end{figure}

Thus we have that the classes GHZ and W are also inequivalent even under most
general LOCC, whereas e.g. a measurement of the projector $P=|+\rangle\langle+|$
with $|+\rangle=1/\sqrt{2}(|0\rangle+|1\rangle)$ in party $A$ maps states within
the classes $W$ (\ref{Wclass}) and $GHZ$ (\ref{GHZclass}) to states within the
class $A-BC$. In a similar way, non--invertible local operators (local, standard measurements) can
convert states within one of the classes $\kappa-\mu\nu$ to states within the
class $A-B-C$. Note that in all cases described above, the inverse
transformations, e.g. from the class $A-B-C$ to one of the classes
$\kappa-\mu\nu$ are impossible as they would imply an increase of the rank of at
least one of the reduced density operators $\rho_A,\rho_B,\rho_C$. These results
are summarized in Fig. \ref{Fig1}.


\subsection{Measures of entanglement and classes under SLOCC}

Several measures have been introduced so far in the literature in order to
quantify entanglement. Although this section is mainly concerned with
qualitative aspects of multipartite quantum correlations, we would like to
relate some of these measures, namely some bipartite ones and the
tripartite 3-tangle \cite{Wo99}(see appendix B), to our classification.
Remarkably, the existence of two kinds of genuine tripartite entanglement in a
three-qubit system, as well as the inequivalence between bipartite and
tripartite entanglement, can be easily understood from the non-increasing
character of these measures under LOCC. In addittion, the 3-tangle allows for a
systematic and practical identification of which class under SLOCC any pure
state belongs to.

For each $\kappa=A,B$ and $C$ we can regard the three-qubit system as a
bipartite system, with qubit $\kappa$, say $A$ for concreteness, being one part
of the system and the remaining two qubits, $B$ and $C$, being the other.
Correspondingly, a state $\ket{\psi}$ of the three qubits can be viewed as a
bipartite state $\ket{\psi_{A(BC)}}$. For bipartite states several measures are
known, which are entanglement monotones \cite{Vi00J}; that is, which cannot be
increased, on average, under LOCC. For instance, we already mentioned the
entropy of entanglement $E(\psi)$ for asymptotic conversions
--given by the entropy $S_A$ of the eigenvalues of
$\rho_A$--- and the monotone $E_2(\psi)$ for
the single copy case ---which is given by the smallest eigenvalue $\lambda_2$ of
$\rho_A$. They all satisfy that vanish for product states (corresponding to
$\rho_A$ with rank 1) while having a positive value for any other state
(corresponding to $\rho_A$ with rank 2). Thus we can interpret the inequivalence
under SLOCC of states whose reduced density matrices differ in rank also in
terms of the impossibility of creating any of the bipartite measures. For
instance, a state in the $A-BC$ class has $S_A=0$, and thus cannot be
transformed with any finite probability into a state of the $AB-C$ class,
because this would have $S_A>0$. We conclude that the monotonicity of these
measures readily split the set of pure states of three qubits into five subsets
which are inequivalent under SLOCC, namely unentagled states $A-B-C$, the three
classes $A-BC$, $AB-C$ and $C-AB$ containing only bipartite entanglement, and a
fifth subset of entangled states with $S_A,S_B,S_C \neq 0$ (i.e.
r$(\rho_A)=$r$(\rho_B)=$r$(\rho_C)=2$). Bipartite measures cannot, however,
determine the inequivalence of the GHZ and W classes.

Is there any known measure of tripartite entanglement which can distinguish between
these two classes? The 3-tangle does. Indeed, it can be computed from the
product decompositions (\ref{GHZclass}) and (\ref{Wclass}) (see \cite{Wo99} for
details), and reads
\be
\tau(\psi_{GHZ})=(2Ks_\alpha s_\beta s_\gamma s_\delta c_\delta)^2 \label{tangle} \not =0\label{tauGHZ}
\ee
for any state in the GHZ class, while it vanishes for any state in the W class.
In the appendix B we prove that the 3-tangle is an entanglement monotone, a very
desirable property for any quantity aiming at measuring entanglement.
Consequently, a state in the W class cannot be transformed by means of LOCC (and
in particular SLOCC) to a state in the GHZ class, which is an independent proof
of the fact that the two kinds of true tripartite entanglement are indeed
inequivalent under SLOCC.


\subsection{Practical identification}

Given an arbitrary state $\ket{\psi}$ of three qubits, expressed in any basis,
it may be interesting to know, for instance, whether it can be converted by
means of LOCC into a GHZ or a W state, if any, or into a EPR state shared
between two of the parties. In our original analysis of the classes we already
have provided a constructive method, based on the analysis of r$(\rho_{\kappa})$
and R$(\rho_{BC})$, to determine the class of $\ket{\psi}$ under SLOCC.
Analysing the R$(\rho_{BC})$ may, however, not be the most practical way to
proceed. Here we suggest to proceed instead according to the following two
steps:

\begin{itemize}
\item compute $\rho_{\kappa}$, $\kappa=A,B$ and $C$, and check whether they have
a vanishing determinant. [note that det$\rho_{\kappa}=0 \Leftrightarrow
S_{\kappa} = 0 \Leftrightarrow $r$(\rho_{\kappa})=1$]
\item If none of the previous determinants vanish [that is, $\ket{\psi}$ has
true tripartite entanglement], then compute the $3$-tangle using the recipe introduced in
\cite{Wo99}.
\end{itemize}
Then Table I, which sumarizes the relation between classes under SLOCC and
measures of entanglement, can be used to catalogue state $\ket{\psi}$.


\section{State $\ket{W}$ and residual bipartite entanglement.}

As mentioned in the previous section, in several aspects the state $\ket{GHZ}$
is the maximally entangled state of three qubits. It also has the feature that
when one of the qubits is traced out, then the remaining two are completely
unentangled. This means, in particular, that if one of the three parties sharing
the system decides not to cooperate with the other two, then they can not use at
all the entanglement resources of the state. The same happens if for some reason
the information about one of the qubits ---namely the identity of the
corresponding states $\ket{0}$ and $\ket{1}$ in (\ref{GHZ})--- is lost.

Here we would like to investigate the robustness of the entanglement of a
three-qubit state $\ket{\psi}$ against disposal of one of the qubits \cite{Br00}. The
residual, two-qubit states $\rho_{AB}$, $\rho_{AC}$ and $\rho_{BC}$ are in
general mixed states. There are several measures of entanglement of mixed states
and therefore multiple ways of quantifying how much (mixed-state) bipartite
entanglement the state $\ket{\psi}$ turns into when one of the qubits is traced
out. Nevertheless, most of the criteria we have examined coincide in pointing
out the state $\ket{W}$ as the one that maximally retains bipartite
entanglement. Note that the reduced density matrix of $|W\rangle$ is
identical for any two subsystems and is e.g. given by
\be
\rho_{AB}=\frac{2}{3}|\Psi^+\rangle\langle\Psi^+|+\frac{1}{3}|00\rangle\langle00|,\label{reducedW}
\ee
with $|\Psi^+\rangle=1/\sqrt{2}(|01\rangle+|10\rangle)$ being a maximally
entangled state of two qubits. Note that one can obtain from a single copy of
(\ref{reducedW}) a state which is arbitrarily close to the state $|\Psi^+\rangle$ by
means of a filtering measurement \cite{Gi96}.


\subsection{Average residual entanglement}

Let us consider first which is the amount of bipartite entanglement, according
to some measure $\E(\rho)$, that the two remaining qubits retain on average when
a third one is traced out, that is,
\be
\bar{\E}(\psi) \equiv \frac{1}{3}(\E(\rho_{AB})+ \E(\rho_{AC}) + \E(\rho_{BC})).
\label{average}
\ee
In general, computing the amount of entanglement $\E(\rho)$ for bipartite mixed
states is a difficult problem. However numerical results have shown that
$\ket{W}$ maximizes the average entanglement of formation, that is the choice
$\E(\rho) = E_f(\rho)$, where $E_f(\rho)$ \footnote{The entanglement of
formation is given by $E_f(\rho)=h(\frac{1}{2}+\frac{1}{2}\sqrt{1-{\cal C}^2})$,
where ${\cal C}$ is the concurrence and $h$ is the binary entropy function
$h(x)=-x{\rm log}_2x-(1-x){\rm log}_2(1-x)$.} is the minimal amount of bipartite
pure-state entanglement [as quantified by means of the entropy of entanglement]
required to prepare locally one single copy of the state $\rho$ \cite{Vi00}.

In addition, we have managed to show analytically (see appendix C) for the
particular choice $\E(\rho) = \c(\rho)^2$, where $\c(\rho)$ is the concurrence
(for a definition of the concurrence see e.g. \cite{Wo99}), the state $\ket{W}$
reaches the maximal average value $\bar{\c^2}(W)=4/9$, which no other
state can match.


\subsection{Least entangled pair}

Another way of quantifying how resistent the entanglement of a tripartite state
$\ket{\psi}$ is to dismissal of one part of the system consists in looking at
the least entangled of the three possible remaining parts, namely at the
function
\be
\E_{\min}(\psi) \equiv \min(\E(\rho_{AB}), \E(\rho_{AC}),  \E(\rho_{BC})).
\label{worstcase}
\ee
For this ``worst case scenario'' we have been able to prove analytically (see
appendix C) that the maximal value of $\E_{\min}(\psi)$ is obtained by the state
$\ket{W}$ for any bipartite measure $\E(\rho)$ which is monotonic with the
concurrence, $\c(\rho)$, such as the entanglement of formation $E_f(\rho)$ and
the monotone $E_2(\rho)$ \footnote{The entanglement monotone $E_2$, expressed in
terms of the concurrence $\c$ is given by
$E_2(\rho)=\frac{1}{2}-\frac{1}{2}\sqrt{1-{\cal C}^2}$.}, which denotes the
minimal amount of bipartite pure-state entanglement [quantified by means of
$E_2(\psi)$] required to prepare locally one single copy of the state $\rho$.

\vspace{.3cm}

We conclude that the state $\ket{W}$ is the state of three-qubits whose
entanglement has the highest degree of endurance against loss of one of the
three qubits. We conceive this property as important in any situation where one
of the three parties sharing the system, say Alice, may suddenly decide not to
cooperate with the other two. Notice that even in the case that Alice would
decide to try to destroy the entanglement between Bob and Claire, this would not
be possible, since any local action on A cannot prevent Bob and Claire from
sharing, at least, the entanglement contained in $\rho_{BC}$ (for instance, by
simply ignoring Alice's actions). Therefore, although essentially tripartite,
the entanglement of the state $\ket{W}$ is also readily bipartite, in contrast
to that of the state $\ket{GHZ}$, which only after some local manipulation can
be brought into a bipartite form.


\section{Generalization to $N$ parties}

In this last section we would like to apply the same techniques to analyze the
entanglement of more general multipartite systems. We will learn that the set of
entangled states is a rather inaccessible jungle for the local explorer, for two
pure states $\ket{\psi}$ and $\ket{\phi}$ are typically not connected by means
of LOCC, so that the parties are usually unable to convert states locally. We
will also study generalizations to $N$ qubits of the state $\ket{W}$.


\subsection{Local inaccessibility of states in general multipartite systems}

Let us consider first $N$ parties each possessing a qubit. The Hilbert space of
the system is
\be
\H^{(N)}=\underbrace{\C^2\otimes\C^2\otimes...\otimes \C^2}_{N},
\ee
and therefore up to a global, physically irrelevant complex constant, a generic
vector depends on $2(2^N-1)$ real parameters. On the other hand we want to
identify vectors which are related by means of a ILO. A
general one-party, invertible operator $A$ must have non-vanishing determinant,
which we can fix to one, det$A=1$, because the operator $kA$ only differs in
that it introduces in the transformed states an extra constant factor $k\in\C$,
which we have already addressed. That is, $A\in SL_{2}(\C)$, and it depends on
$6$ real parameters. Therefore the set of equivalence classes under SLOCC,
\be
\frac{\H^{(N)}}{\underbrace{SL_{2}(\C)\times SL_{2}(\C) \times ... \times SL_{2}(\C)}_{N}},
\ee
depends {\em at least} on $2(2^N-1) - 6N$ parameters. This lower bound allows
for a finite number of classes for $N=3$, but shows that for any larger number
$N$ of qubits there are infinitely many classes, labeled by at least one
continuous parameter. The reason is that the number of parameters from a state
$\ket{\psi}$ which the parties can modify by means of a general ILO $A\otimes
B\otimes...\otimes N$ grows linearly with $N$ ($6N$ for the multi-qubit case),
whereas the number of parameters required to specify $\ket{\psi}$ grows
exponentially with $N$.

More generally, if the Hilbert space is given by $\H=\C^{n_1}\otimes ... \otimes
\C^{n_N}$, then the set of equivalence classes under SLOCC,
\be
\frac{\C^{n_1}\otimes ... \otimes \C^{n_N}}{SL_{n_1}(\C)\times...\times SL_{n_N}(\C)},
\ee
depends at least on $2(n_1n_2...n_N-1) - 2\sum_{i=1}^N(n_i^2-1))$. This shows
that only for $N=3$ there are still some systems with (potentially) only a
finite number of classe under SLOCC, namely those with Hilbert space
$\C^2\otimes\C^{n_2}\otimes\C^{n_3}$, that is, having a qubit as at least one of
the subsystems. In all other cases, one finds an infinite number of classes.

We notice that even allowing for non-invertible local operations the amount of
parameters that can be changed by local manipulations is typically smaller than
that the state depends on. That is, the subset of states that can be reached
locally from a given state $\ket{\psi}$ is of zero measure in the set of states
of the multipartite system. Recall that in the bipartite scenario,
$\H=\C^n\otimes\C^m$, there is always a maximally entangled state from which all
the other states can be locally prepared with certainty of success. We see now
that, in constrast, there is typically in a multipartite system no state from
which all the others can be prepared, not even with some probability of success. Of
course, the parties can always resort to, say, using a sufficient amount of EPR
states distributed among them to prepare any multipartite state by standard
teleportation. This implies, however, using an initial state (that of many EPR
states) which belongs to a Hilbert space much larger than the Hilbert space of
the state the parties are trying to create, and thus does not change the previous
conclusion.


\subsection{State \ket{W} in multi-qubit systems}

Let us have a look at the generalized form $|W_N\rangle$ of the state
$|W\rangle$ (\ref{W}). We define the state
\be
|W_N\rangle \equiv 1/\sqrt{N}|N-1,1\rangle,
\ee
where $|N-1,1\rangle$ denotes the totally symmetric state including $N-1$ zeros
and $1$ ones. For example, we obtain for $N=4$
\be
|W_4\rangle=1/\sqrt{4}(|0001\rangle+|0010\rangle+|0100\rangle+|1000\rangle).
\ee
One immediately observes that the entanglement of this state is again very
robust against particle losses, i.e. the state $|W_N\rangle$ remains entangled
even if any $N\!-\!2$ parties lose the information about their particle. This means
that any two out of $N$ parties possess an entangled state, independently of
whether the remaining $(N-2)$ parties decide to cooperate with them or not. This
can be seen by computing the reduced density operator $\rho_{AB}$ of
$|W_N\rangle$, i.e. by tracing out all but the first and the second systems. By
symmetry of the state $|W_N\rangle$, we have that all reduced density operators
$\rho_{\kappa\mu}$ are identical and we obtain
\be
\rho_{\kappa\mu}=\frac{1}{N}(2 |\Psi^+\rangle\langle\Psi^+|+(N-2)|00\rangle\langle 00|).
\ee
The concurrence can easily determined to be
\be
{\cal C}_{\kappa\mu}(W_N)=\frac{2}{N},
\ee
which shows that $\rho_{\kappa\mu}$ is entangled, even distillable. We
conjecture that the average value of the square of the concurrence for $\ket{W_N}$,
\be
\frac{2}{N(N-1)}\sum_{\kappa} \sum_{\mu\neq\kappa} {\cal C}^2_{\kappa\mu}(W_N)=\frac{4}{N^2},
\ee
is again the maximal value achievable for any state of $N$ qubits.

\section{Summary and conclusions}

In this work, we investigated equivalence classes of multipartite states
specified by stochastic local operations and classical communication. We showed
that for pure states of three qubits there are 6 different classes of this kind.
In particular, we found that there are two inequivalent types of genuine
tripartite entanglement, represented by the GHZ state and the state W. We showed
that the state W is the state of three qubits that retains a maximal amount of
bipartite entanglement when any one of the three qubits is traced out. For
multipartite ($N\geq 4$) and multilevel systems, we showed that there exist
infinitely many inequivalent kinds of entanglement (i.e. classes under SLOCC).

\section*{Acknowledgments}
This work was supported by the Austrian Science Foundation under the SFB
``control and measurement of coherent quantum systems'' (Project 11), the
European Community under the TMR network ERB--FMRX--CT96--0087, the European
Science Foundation and the Institute for Quantum Information GmbH. G.V also
acknowledges a Marie Curie Fellowship HPMF-CT-1999-00200 (European Community).

\section*{Appendix A: SLOCC and local ranks}

In this appendix we show that states $\ket{\psi}$ and $\ket{\phi}$ belong to the
same class under SLOCC iff they are related by means of a invertible local
operator (ILO). From this connection it follows easily that the local ranks of a pure
state, r($\rho_{\kappa})$, $\kappa=A,B,...$, are invariant under SLOCC, whereas
under LOCC they can only decrease.

{\bf Lemma:} If the bipartite vectors $\ket{\psi}$ and $\ket{\phi} \in \C^n\otimes\C^m$ fulfill
\be
\ket{\phi} = A\otimes 1_B \ket{\psi},
\ee
then the ranks of the corresponding reduced density matrices satisfy
r$(\rho_A^{\psi}) \geq $ r$(\rho_A^{\phi})$ and  r$(\rho_B^{\psi}) \geq
$r$(\rho_B^{\phi})$.

{\bf Proof:} We consider the Schmidt decomposition of $\ket{\psi}$,
\be
\ket{\psi} = \sum_{i=1}^{n_{\psi}} \sqrt{\lambda_i^{\psi}}\ket{i}\ket{i},~~~ \lambda^{\psi}_i > 0,~~ n_{\psi}\leq \min (n,m),
\ee
and write the operator $A$ as
\be
A = \sum_{i=1}^n \ket{\mu_i}\bra{i},
\label{operator}
\ee
where $\ket{\mu_i} \in \C^n$ do not need to be normalized nor linearly
independent. Then we have that $\rho_A^{\psi} =
\sum_{i=1}^{n_{\psi}}\proj{i}$ and $\rho_A^{\phi} = A\rho_A^{\psi}A^{\dagger} =
\sum_{i=1}^{n_{\psi}}\proj{\mu_i}$, so that r$(\rho_A^{\phi})\leq n_{\psi}$. The
second inequality of the Lemma follows from the fact that for any bipartite vector
r$(\rho_A)=$ r$(\rho_B)$. $\Box$

{\bf Corollary:} If the vectors
$\ket{\psi},\ket{\phi}\in\H_A\otimes\H_B\otimes...\otimes\H_N$ are connected by
a local operator as $\ket{\phi}=A\otimes B\otimes...\otimes N \ket{\psi}$, then
the local ranks satisfy r$(\rho_{\kappa}^{\psi}) \geq $
r$(\rho_{\kappa}^{\phi})$, $\kappa=A, B,..., N$.

{\bf Proof:} Indeed, for each of the parties, say Alice for concreteness, we can
view the operator $A\otimes B\otimes...\otimes N$ as the composition of two local operators, $A\otimes
1_{B...N}$ and $1_A\otimes (B\otimes...\otimes N)$, and the Hilbert space as
$\H_A\otimes \H_{B...N}$. Then, because of the previous lemma, application of
the first operator cannot increase r($\rho_A$), and the same happens with the
second operator, which cannot increase  r$(\rho_{B...N})$ [recall that for any
pure state r$(\rho_A)=$ r$(\rho_{B...N})$]. $\Box$

{\bf Theorem:} Two pure states of a multipartite system are equivalent under
SLOCC iff they are related by a local invertible operator.

{\bf Proof:} If
\be
\ket{\phi}=A\otimes B\otimes...\otimes N \ket{\psi},
\label{localoperator}
\ee
then a local protocol exists for the parties to transform $\ket{\psi}$ into
$\ket{\phi}$ with a finite probability of success. Indeed, each party needs
simply perform a local POVM including a normalized version of the corresponding
local operator in (\ref{localoperator}). For instance, Alice has to apply a POVM
defined by operators $\sqrt{p_A}A$ and $\sqrt{1_A - p_AA^{\dagger}A}$, where
$p_A\leq 1$ is a positive weight such that $p_AA^{\dagger}A \leq 1_A$, and
similarly for the rest of the parties. Then such a local protocol converts
$\ket{\psi}$ succesfully into $\ket{\phi}$ with probability $p_Ap_B...p_N$. If,
in addition, $A,B,...,N$ are invertible operators, then obviously
\be
\ket{\psi}=A^{-1}\otimes B^{-1}\otimes...\otimes N^{-1} \ket{\phi}
\ee
and the conversion can be reversed locally. Let us then move to prove the
converse. We already argued (section II.A) that if $\ket{\psi}$ can be converted
into $\ket{\phi}$ by LOCC, then a local operator relate them. We want to prove
now that equivalence of $\ket{\psi}$ and $\ket{\phi}$ under SLOCC implies that
this operator can always be chosen to be invertible. For simplicity, we will
assume that $\ket{\psi}$ and $\ket{\phi}$ are related by a local operator acting
non-trivially only in Alice's part,
\be
\ket{\phi} = A\otimes 1_{B...N} \ket{\psi}.
\label{onlyA}
\ee
[The general case would correspond to composing operator $A\otimes 1_{B...N}$
with operator $1_A\otimes B\otimes 1_{C...N}$, and similarly for the rest of the
parties. The following argumentation should then be applied sequentially to each
party individually.] We can then consider the Schmidt decomposition of the
states with respect to part $A$ and part $B...N$
\bea
\ket{\psi} = \sum_{i=1}^{n_{\psi}} \sqrt{\lambda_i^{\psi}}\ket{i}\ket{\tau_i},~~~~~~~~~~~ \lambda^{\psi}_i > 0\\
\ket{\phi} = \sum_{i=1}^{n_{\phi}} \sqrt{\lambda_i^{\phi}}(U_A\ket{i})\ket{\tau_i},~~~~~ \lambda^{\phi}_i > 0
\eea
where the local unitary $U_A$ relate the two local Schmidt basis in Alice's
part, $\{\ket{i}\}_{i=1}^n \in \H_A = \C^n$, $\ket{\tau_i}\in
\H_B\otimes...\otimes\H_N$, and $n_{\psi}=n_{\phi}$ because of the previous
corollary. Now, operator $A$ in equation (\ref{onlyA}) must be of the form (up
to some irrelevant permutations in the Schmidt basis)
\bea
A = U_A (A_1 + A_2)\nonumber\\
A_1 \equiv \sum_{i=1}^{n_{\psi}} \sqrt{\frac{\lambda_i^{\phi}}{\lambda_i^{\psi}}} \proj{i},\\
A_2 \equiv \sum_{i=n_{\psi}+1}^n\ket{\mu_i}\bra{i}
\eea
where $\ket{\mu_i}$ are arbitrary unnormalized vectors. Notice that vectors
$\ket{\mu_i}$ play no role in equation (\ref{onlyA}) since
$A_2\otimes1_{B...N}\ket{\psi}=0$. Therefore we can redefine
\be
A_2 \equiv \sum_{i=n_{\psi} +1}^n \proj{i},
\ee
which implies that $A$ is an invertible operator.$\Box$


\section*{Appendix B: $\tau$ is an entanglement monotone}
In this appendix, we show that the 3-tangle $\tau$ is an entanglement monotone,
i.e. decreasing on average under LOCC in all the three parties. We first note
that any local protocol can be decomposed into POVM's such that only one
party performs operations on the system. This, together with the invariance of the 3-tangle
$\tau$ under permutations of the parties, ensures that it is sufficient to consider a
local POVM in $A$ only. Furthermore, we can restrict ourselves to two--outcome
POVM's due to the fact that a genarlized (local) POVM can be implemented by a
sequence of two outcome POVM's. Let $A_1,A_2$ be the two POVM elements such that
$A_1^\dagger A_1+A_2^\dagger A_2 = \eins$. We can write $A_i=U_iD_iV$, where
$U_i$, $V$ are unitary matrices and $D_i$ are diagonal matrices with entries
$(a,b)$ $[((1-a^2)^{\frac{1}{2}},(1-b^2)^{\frac{1}{2}}$)] respectively. Note
that we used the singular value decomposition for $A_i$, and we have that the restriction
that $A_1,A_2$ constitute a POVM immediately implies that the unitary operation
$V$ can be chosen to be the same in both cases. We consider an initial state
$|\psi\rangle$ with 3-tangle $\tau(\psi)$. Let $|\tilde\phi_i\rangle=
A_i|\psi\rangle$ be the (unnormalized) states after the application of the POVM.
Normalizing them, we obtain $|\phi_i\rangle=|\tilde\phi_i\rangle/\sqrt{p_i}$
with $p_i=\langle\tilde\phi_i|\tilde\phi_i\rangle$ and $p_1+p_2=1$. We want to
show that $\tau^\eta$, $0<\eta\leq1$ is, on average, always decreasing and thus
an entanglement monotone, i.e for
\be
<\tau^\eta>= p_1\tau^\eta(\phi_1)+ p_2 \tau^\eta(\phi_2)\label{monotontau}
\ee
we have that
\be
<\tau^\eta> \leq \tau^\eta(\psi) \label{monoton}
\ee
is fulfilled for all possible choices of the POVM $\{A_1,A_2\}$. Using that
$\tau$ is invariant under local unitaries, we do not have to consider the
unitary operations $U_i$ in our calculations, i.e.
$\tau(U_iD_iV\psi)=\tau(D_iV\psi)$. Taking this simplification into account, a
straightforward calculation shows that
\be
\tau(\phi_1)=\frac{a^2b^2}{p_1^2} \tau(\psi) , \mbox{  } \tau(\phi_2)=\frac{(1-a^2)(1-b^2)}{p_2^2} \tau(\psi),
\ee
where we used that $\tau(\epsilon \tilde{\phi_i})=\epsilon^4
\tau(\tilde{\phi_i})$, which can be checked by noting that $\tau$ is a quartic
function with respect to its coefficients in the standard basis\cite{Wo99}. Note that
the dependence of $\tau(\phi_i)$ on the unitary operation $V$ is hidden in
$p_i$. For $\eta=1/2$, one obtains for example
$\tau^{\frac{1}{2}}(\phi_1)=ab/p_1\tau^{\frac{1}{2}}(\psi)$. Substituting in
(\ref{monotontau}), we find
\be
<\tau^{\frac{1}{2}}>=(ab+\sqrt{(1-a^2)(1-b^2)}) \tau^{\frac{1}{2}}(\psi). \label{tau12}
\ee
In this case, one can easily check that ($\ref{tau12}) \leq \tau^{\frac{1}{2}}$
by noting that (\ref{tau12}) is maximized for $a=b$. We thus have that
$\tau^{\frac{1}{2}}$ is, on average, always decreasing and thus an entanglement
monotone. In a similar way, one can check for $0< \eta \leq 1$ that $\tau^\eta$
is an entanglement monotone. However, for $\eta \not= 1/2$, the derivation is a bit more
involved due to the fact that in this case the propabilities $p_i$ in the
expression for $<\tau^\eta>$ do no longer cancel and have to be calculated
explicitly.


\section*{Appendix C: $|W\rangle$ maximizes residual bipartite entanglement}

Here we show that for all tripartite pure states, except the state $|W\rangle$
the following inequality holds
\be
E_\tau \equiv {\cal C}^2_{AB}+{\cal C}^2_{AC}+{\cal C}^2_{BC} < \frac{4}{3},\label{inequ}
\ee
while the state $|W\rangle$ reaches the value $E_\tau=4/3$. Note that we used the
shorthand notation ${\cal C}_{AB}$ for the concurrence of the reduced density
operator $\rho_{AB}, {\cal C}(\rho_{AB})$, and similary for  ${\cal
C}_{AC}$,${\cal C}_{BC}$.

Inequality (\ref{inequ}) already implies that the state $|W\rangle$ reaches the
maximum average value $\bar{\E}(\psi)$ of Equ. (\ref{average}) for the choice of $\E(\rho) =
\c(\rho)^2$, namely $\bar{\E}(W)=4/9$.

At the same time, inequality (\ref{inequ}) also shows that the state $|W\rangle$
maximizes the function $\E_{\min}(\psi)$ (\ref{worstcase}) for the choice of
$\E(\rho) = \c(\rho)^2$, since (\ref{inequ}) implies that
\be
{\cal C}^2_{\min}(\psi) \equiv \min({\cal C}^2_{AB},{\cal C}^2_{AC},{\cal C}^2_{BC}) < 4/9 \label{mini}
\ee
for all states except the state $|W\rangle$, for which the value $4/9$ is
reached. From (\ref{mini}) follows that for any bipartite measure of
entanglement $\E(\rho)$ which is monotonically increasing with the square of the
concurrence (and hence with the concurrence itself), the state $|W\rangle$
maximizes the function $\E_{\min}(\psi)$ (\ref{worstcase}), i.e.
\be
\E_{\min}(\psi) < \E_{\min}(W) = \E({\cal C}^2=4/9).
\ee
Assume that this is not the case, i.e. there exist a state $\psi$ for which
$\E_{\min}(\psi) > \E_{\min}(W)$. Since by assumption $\E$ is monotonically
increasing with the concurrence, this would imply that ${\cal C}^2_{\min}(\psi)
>4/9$, which contradicts Equ. (\ref{mini}) and is hence impossible.

Note in addition that any good measure of entanglement should be a convex
function \cite{Vi00J}, as $\c(\rho), E_f(\rho)$ and $E_2(\rho)$ are. This
implies, when applied to (\ref{average}) and (\ref{worstcase}) that the optimal
values for $\bar{\E}$ and $\E_{\min}$ are achieved for pure states.

Ther remainder of this appendix is devoted to prove inequality (\ref{inequ}). Using
the definition of the 3-tangle, $\tau\equiv\tau_{ABC}={\cal C}^2_{A(BC)}-{\cal
C}^2_{AB}-{\cal C}^2_{AC}$ ~\cite{Wo99}and the invariance of the 3-tangle under permutations
of the parties, we can rewrite $E_\tau$ as $1/2({\cal C}^2_{A(BC)}+{\cal
C}^2_{B(AC)}+{\cal C}^2_{C(AB)}-3\tau)$. Using that ${\cal
C}^2_{\kappa(\mu\nu)}=4{\rm det}\rho_{\kappa}$, we can evaluate $E_\tau$ for the
different classes.

Starting with the class $A-B-C$, we immeadetly obtain that
$E_\tau(\Psi_{A-B-C})=0$. For the class $A-BC$, we have that $\tau=0$ and
${\cal C}^2_{A(BC)}=0$. Since ${\cal C}^2_{B(AC)},{\cal C}^2_{C(AB)} \leq 1$, we have that
$E_\tau(\Psi_{A-BC}) \leq 1$ in this case (and similary for the classes $B-AC,
C-AB$).

Now we consider the class $W$, specified by equ. (\ref{Wclass}). Again,
we have that $\tau=0$. We find that $E_\tau(\Psi_W)=4(ab+ac+bc)$ (which does not
depend on $d$). Notice that $E_\tau$ is maximized for $a=b=c=1/3$ - which
corresponds to the state $|W\rangle$ - and leads to $E_\tau=4/3$. For all other
values of $a,b,c,d$, we have that $E_\tau < 4/3$.

Let us now turn to the class GHZ, specified in eq. (\ref{GHZclass}). Using that
$\tau(\Psi_{GHZ})$ is given in eq. (\ref{tauGHZ}) and
$\det\rho_A=K^2c_\delta^2s_\delta^2s_\alpha^2(1-c_\beta^2c_\gamma^2)$ (and
similary for $\det\rho_{B,C}$), we obtain
\be
E_\tau=\frac{4c_\delta^2s_\delta^2[(s_\alpha^2s_\beta^2+s_\alpha^2s_\gamma^2+s_\beta^2s_\gamma^2)-3s_\alpha^2s_\beta^2s_\gamma^2]}{(1+2c_\delta s_\delta c_\alpha c_\beta c_\gamma c_\varphi)^2} \label{Etau}
\ee
One readily checks that (\ref{Etau}) is maximized for $\delta=\pi/4$ and
$\varphi=\pi$ (which corresponds to $c_\delta=s_\delta=1/\sqrt{2}$ and
$c_\varphi=-1$), independent of the values of $\alpha,\beta,\gamma \in
(0,\pi/2]$. Thus we have that $E_\tau \leq E_\tau(\delta=\pi/4,\varphi=\pi)$ and
after some algebra we obtain
\be
E_\tau \leq \frac{(c_\alpha^2+c_\beta^2+c_\gamma^2)-2(c_\alpha^2c_\beta^2+c_\alpha^2c_\gamma^2+c_\beta^2c_\gamma^2)+3c_\alpha^2c_\beta^2c_\gamma^2}{(1+c_\alpha c_\beta c_\gamma )^2} \label{Etau1}
\ee
We want to show that the (rhs) of eq. \ref{Etau1} $< 4/3$. Let us call $x\equiv
c_\alpha,y\equiv c_\beta,z\equiv c_\gamma$ with $0\leq x,y,z<1$. We thus have to show that
\bea
f(x,y,z)\equiv&&3(x^2+y^2+z^2)-6(x^2y^2+x^2z^2+y^2z^2) \nonumber \\
&+&5(x^2y^2z^2)-4+8xyz < 0
\eea
Let us calculate the maximum of $f(x,y,z)$. We therefore take the derivatives of
$f(x,y,z)$ with respect to $x,y,z$ respectively (which we denote by
$f_x,f_y,f_z$) and set them to zero. One immeadetly observes (by considering
linear combination of the resulting equations, e.g. $xf_x-yf_y$, where one e.g.
obtains $(x^2-y^2)(1-2z^2)=0$), that for a maximum we must have $x=y=z$. The
possible solutions of the resulting polynomial of degree 5 can be checked to lie
outside the intervall $[0,1)$, i.e. outside the range of $x,y,z$ except for
$x=y=z=0$. It can however be easily verified that this solution give rise to a
minimum of $f(x,y,z)$, namely $f(0,0,0)=-4$. Thus the maximum of $f(x,y,z)$ is
obtained at the border of the range for $x,y,z$, which corresponds to the
surfaces of a cube. Due to the fact that $f(x,y,z)$ is invariant under
permutations of the variables, we only have to check two of the surfaces, e.g.
the surfaces specified by $x=0$ and $x=1$ (actually $x=1-\epsilon$, where
$\epsilon$ is an infinitesimally small positive number) and we find (i)
$f(0,x,y)=3(y^2+z^2)-6y^2z^2-4 \leq -1$ (the maximum in this case is e.g.
obtained for $y=0,z=1-\epsilon$)) and (ii) $f(1,y,z)=8yz-3(y^2+z^2)-y^2z^2-1 <
0$. In (ii), it can be checked that a necessary condition for a maximum is $y=z$
and that $f(1,y,y)$ is monotonically increasing in $[0,1)$ and is thus maximized
for $y=z=(1-\epsilon)$. One obtains $f(x,y,z) \leq f(1,1-\epsilon,1-\epsilon) <
0$ as desired.

So we managed to show that the state $|W\rangle$ is the only state which
fulfills $E_\tau=4/3$, and for all other tripartite pure states we have that
$E_\tau < 4/3$.


 \vspace{5cm}

\narrowtext
\begin{table}
\begin{tabular}[t]{||c|l|l|l|l||}
Class   & $S_A$ & $S_B$ & $S_C$ & $\tau$  \\ \hline
A-B-C   & 0 & 0 & 0 & 0     \\ \hline
A-BC    & 0 & $>0$  & $>0$  & 0     \\ \hline
B-AC    & $>0$  & 0 & $>0$  & 0     \\ \hline
C-AB    & $>0$  & $>0$  & 0 & 0     \\ \hline
W   & $>0$  & $>0$  & $>0$  & 0     \\ \hline
GHZ & $>0$  & $>0$  & $>0$  & $>0$     \\
\end{tabular}
\caption[]{Values of the local entropies $S_A, S_B, S_C$ and the 3-tangle $\tau$ for the different classes}
\label{Table1}
\end{table}


\begin{references}
\bibitem{comment}
Different classifications based on LU invariants were recently proposed in
\cite{Ac00} and \cite{Ca00}. Depending on the values of these invariants,
several classes of states in one case and types of entanglement in the other are
identified. We want to remark that our classification is based on probabilistic
conversions under LOCC, and therefore, as expectable, our results are not fully
compatible with those of these approaches.
\bibitem{monotones} Recall that the $n$ entanglement monotones
$E_k(\psi)\equiv\sum_{i=k}^n \lambda_i$, where $\lambda_i$ come from the Schmidt
decomposition of $\ket{\psi} \in \C^n\otimes\C^n$,
\be
\ket{\psi} = \sum_{i=1}^n \sqrt{\lambda_i} \ket{i}\otimes\ket{i},~~ \lambda_i\geq\lambda_{i+1}\geq 0,
\ee
are the measures that provide a quantitative description of the entanglement
resources of a single copy of pure states, in that, for instance, they give the
optimal probability $P(\psi\rightarrow\phi)$ of conversion of the state
$\ket{\psi}$ into the state $\ket{\phi}$ under LOCC \cite{Vi99}, namely
\be
P(\psi\rightarrow\phi)= \min \{\frac{E_1(\phi)}{E_1(\psi)},...,\frac{E_n(\phi)}{E_n(\psi)}\}.
\ee
In a two-qubit system we have only two non-trivial monotones,
$E_1(\psi)=\lambda_1+\lambda_2 =1$ and $E_2=\lambda_2\leq 1/2.$ In a similar way
as $E_2(\psi)>E_2(\phi)$ determines that $\ket{\psi}$ can be transformed into
$\ket{\phi}$ with certainty by means of LOCC \cite{Ni99}, the extension to mixed
states $\rho$ of two qubits, $E_2(\rho)$, also says when the entanglement of the
pure state $\ket{\psi}$ suffices to locally prepare $\rho$ with certainty
\cite{Vi00}.
\bibitem{product}
A state $t\ket{00}+x\ket{01}+y\ket{10}+z\ket{11}$ is product
iff it can be written as $(a\ket{0}+b\ket{1})(c\ket{0}+d\ket{1})$, that is iff
tz=xy. A complex linear combination
$\lambda_1\ket{00}+\lambda_2(x\ket{01}+y\ket{10}+z\ket{11})$ with $\lambda_2\neq
0$ cannot be product for any $\lambda_1$ iff $z=0$.
\end{references}
\end{document}